# On the Onset of the Flash Transition in Single Crystals of Cubic Zirconia

Yanhao Dong (dongya@seas.upenn.edu)

*Department of Materials Science and Engineering, University of Pennsylvania,*

*Philadelphia, PA 19104, USA*

**Abstract**

The reported onset temperatures of the flash transition in cubic zirconia single crystals have been analyzed in the present note. The analysis for high-temperature low-field data gives an activation energy of 0.99 eV for the charge transport, which agrees well with the measured ionic conductivity data in cubic zirconia single crystal. Electrical reduction is believed to play an additional role under low-temperature high-field conditions, facilitating electrical/thermal runaway under more modest conditions.

**I. Introduction**

Flash sintering[1-3] as an emerging sintering technique has attracted great attention in recent years. In a standard flash sintering experiment, a ceramic powder compact is placed under a constant voltage in a ramping furnace. A sudden electrical and thermal runaway takes place when the furnace reaches a critical temperature, leading to a rapid densification that completes sintering in a few seconds. While the mechanism responsible for the rapid mass transport is still controversial and under debate, thermal runaway is of central importance as argued by several groups[4-9]. In a recent paper by Yadav and Raj[10], the authors reported onset temperature of flash transition in 8 mol% yttria stabilized zirconia single crystals as a function of the applied electric field and obtained an apparent activation energy of 0.66 eV (64 kJ/mol) for an unknown charge transport process. If thermal runaway is responsible for the onset of flash sintering, now in single crystals without the complications of the porous powder compact and any sintering effect, one should be able to obtain the correct activation energy as claimed in Ref. 5. In the present note, we shall analyze the onset data reported by Yadav and Raj, and hopefully provide a better interpretation.

## II. Data analysis and discussion

According the analysis in Ref. 5, if plotting $\ln(E^2/T_{on}^4)$ vs. $1/T_{on}$, one obtains a slope of $E_a/k_B$, where $E$ is the applied electric field, $T_{on}$ is the onset furnace temperature, $E_a$ is the activation energy of the sample's conductivity and $k_B$ is the Boltzmann constant. Following the same treatment, we re-plot the data in Ref. 10 in **Fig. 1**. Apparently, the linearity is acceptable yet not very good. So we expect additional effects to play a role, i.e. reduction effects as also pointed out by Yadav and Raj. A close inspection of the data shows that the high-temperature low-field data follow a linear dependence well. Hence, we linearly fit these data (the first eight data points in **Fig. 1**, temperatures ranged from 400 to 700 ºC; fitting results shown by the dash line), which gives an activation energy of 0.99 eV. It agrees well with the conductivity data measured in an 8YSZ single crystal (<100> orientation; MTI Corporation, Richmond, CA), which gives an activation energy of 1.00 eV in the same temperature range. Therefore, we conclude the high-temperature low-field data can be totally explained by a thermal runaway model.

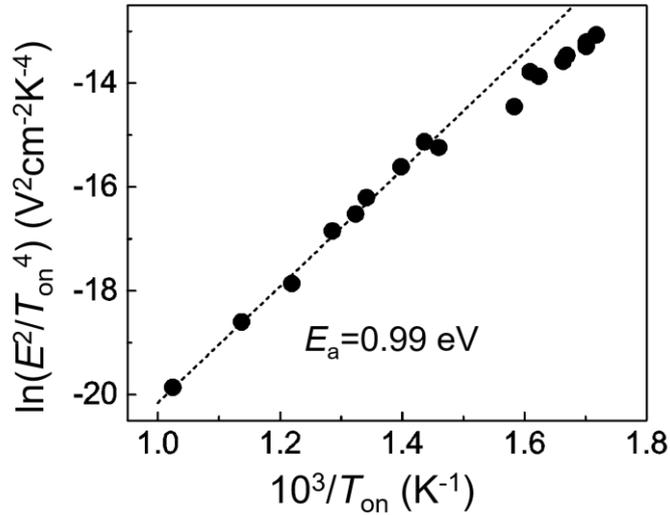

**Figure 1** Onset data from Ref. 10 on $\ln(E^2/T_{on}^4)$ vs. $T_{on}^{-1}$ plot with the slope indicating $E_a/k_B$.

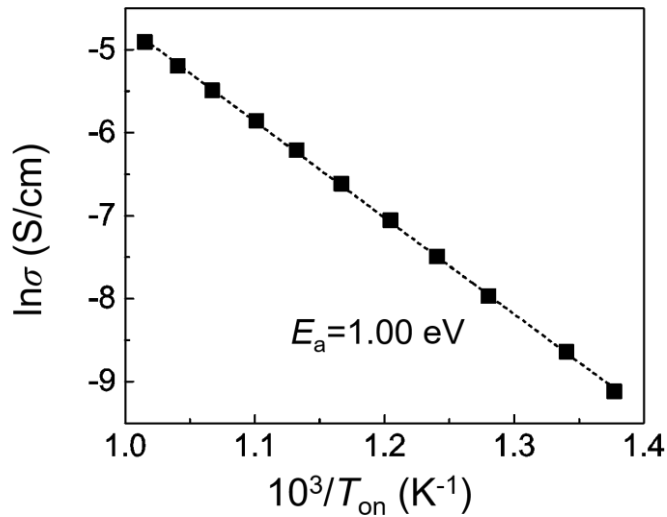

**Figure 2** Conductivity of 8YSZ single crystal in the temperature range of 450 to 700 ºC. Measurements were conducted using AC impedance method to exclude the electrode resistance.

As to the low-temperature high-field data, reduction is expected to happen and electronic conduction can arise. This would lead to a decrease of the onset furnace temperature if under the same field or a decrease of the required field if under the same onset furnace temperature, which can be qualitatively seen in **Fig. 1**. We have conducted some electrical reduction experiments in 8YSZ single crystal (unpublished results; see below), where to avoid Joule heating we began with a constant, low furnace temperature and imposed a constant voltage across an 8YSZ single crystal. The *I-V* curve as well as the Joule heating power is shown in **Fig. 3** as a function of time. With a minimal Joule heating on the order of 1 mW/mm$^3$ and no ramping in the furnace temperature or the applied voltage, we observed an electrical runaway after a long stagnation of about 26 h. After the electrical runaway, we also observed blackening of the crystal as the evidence for electrical reduction (see various examples in **Fig. 4**). Such an electrical degradation shares the same feature as the well-known resistive degradation in dielectric materials[11,12] and can contribute to the onset of the flash transition in flash sintering. With the results we reported easier where we show electrical reduction can greatly enhance grain growth[13], we see that electrical reduction can influence both the charge (e.g. conductivity and onset of flash transition) and mass transport (e.g. grain growth) in YSZ, thus being of vital importance in understanding the material's behavior under electrical field.

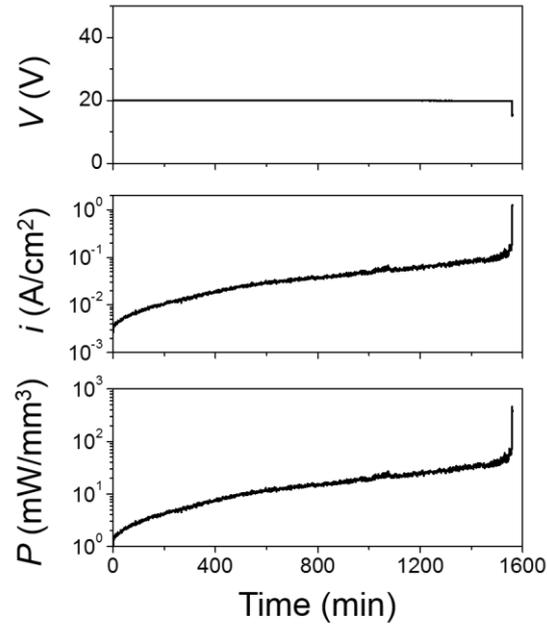

**Figure 3** Sample's voltage $V$, current density $i$ and Joule heating power $P$ trajectories under a furnace temperature of 380 ºC and a total voltage of 20 V. The compliance current is set to be 1.25 A/cm$^2$. Sample thickness: 0.5 mm.

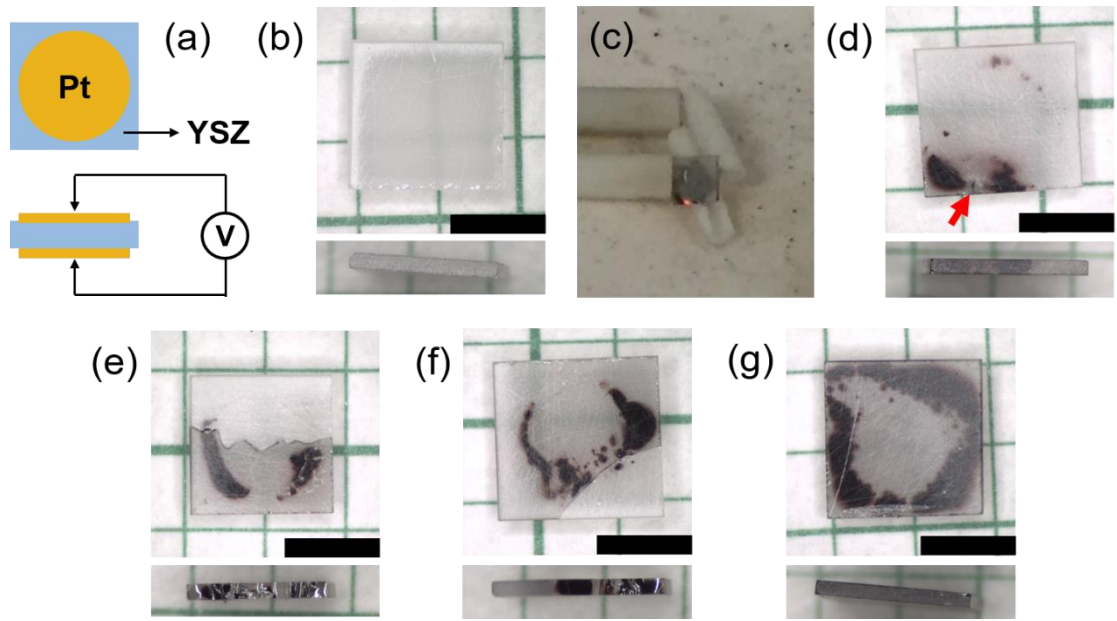

**Figure 4** (a) Schematics for the electrode geometry. Optical images of (b) a blank sample without electrical testing and YSZ single crystals blackened at (d) 380 ºC and 45 V, (e-f) 380 ºC and 30 V and (g) 480 ºC and 15 V. All scale bars are 3 mm. Shown in (c) is the optical image of the degraded sample after electrical runaway and cooled down to room temperature with current running through. The temperature of the hot spot measured by pyrometer is about 850 °C.

**III. Conclusions**

To conclude, we analyzed the reported onset temperatures of the flash transition in cubic zirconia single crystals, following the same method described in Ref. 5. The analysis for high-temperature low-field data gives an activation energy of 0.99 eV for the charge transport, which agrees well with the measured ionic conductivity data in the same single-crystalline sample. Electrical reduction is believed to play an additional role under low-temperature high-field conditions, facilitating electrical/thermal runaway under more modest conditions.

(2016), DOI: 10.1111/jace.14615.